\documentstyle[11pt]{article}
\newtheorem{theorem}{Theorem}[section]
\newtheorem{lemma}[theorem]{Lemma}

\title{The Analytic Quantum Information Manifold}
\author{R. F. Streater,\\Dept. of Mathematics, King's College London,\\
Strand, WC2R 2LS}
\begin{document}
\maketitle
\begin{abstract}
Let $H\geq I$ be a self-adjoint operator on a Hilbert space, such that
$e^{-\beta H}$ is of trace-class for some $\beta<1$. Let
$V$ be a symmetric operator such that $\|V\|_\omega:=\|RV\|<\infty$, where
$R=H^{-1}$. We show that the partition function ${\rm Tr}\,e^{-(H+\lambda V)}$
is analytic in $\lambda$ in a hood of the origin in the sense of
Fr\'{e}chet, in the Banach space with norm $\|\bullet\|_\omega$.
This is applied to the quantum information manifold defined by $H$.
\end{abstract}
\section{Introduction}
The information manifold in non-parametric estimation theory has been
introduced and studied by Pistone and Sempi \cite{Pistone,Gibilisco}.
For a finite number of parameters, the theory is described in the books
\cite{Amari,Kass}. The quantum version has been introduced by Hasagawa
\cite{Hasagawa,Hasagawa2} and by
Nagaoka \cite{Nagaoka}, and by Petz \cite{Petz,Petz2}. This study
is strictly valid only if the Hilbert space ${\cal H}$ of states is finite
dimensional. It provides a geometrical picture, as well as rigour, to the
theory of linear response
\cite{Kubo,Mori,Bogoliubov,Matsubara,Balian,Roepstorff}. For
${\rm dim}{\cal H}=\infty$, but limited to bounded potentials, the theory has
been developed by Araki \cite{Araki} in the context of Tomita-Takesaki
theory. Our more concrete Hilbert-space version of this goes as follows.

Denote by ${\cal C}_1$ the set of trace-class operators on ${\cal H}$ and
let $\Sigma\subseteq{\cal C}_1$ denote the set of density operators. Given
$\rho_0\in\Sigma$ let $H_0=-\log\rho_0+cI\geq I$ be a self-adjoint operator
with domain ${\cal D}(H_0)$ such that
\begin{equation}
\rho_0=Z_0^{-1}e^{-H_0}.
\end{equation}
We recognise $H_0$, which is unique up to the constant $c$, as the
Hamiltonian, and $Z_0$ as the partition function, whose finiteness
expresses thermodynamic stability. Let $X$ be a symmetric bounded operator,
so $X\in{\cal B}({\cal H})$, and perturb $H_0$ to $H_X=H_0+X$, with
corresponding state $\rho_X=Z_X^{-1}\exp(-H_X)$ and free energy
$\psi_X=\log Z_X$. 
Araki proves that $\psi_X$ is an analytic functional
with a convergent Taylor series, whose first few terms are
\begin{eqnarray}
\nabla_Y^+\psi_X&=&\rho_X.Y:=\mbox{Tr}(\rho_XY)\\
\nabla_Y^+\nabla_Z^+\psi_X&=&\int_0^1d\lambda\mbox{Tr}\left\{\rho_X^\lambda
\hat{Y}\rho_X^{\lambda^\prime}\hat{Z}\right\}:=g_X(Y,Z)\label{met}\\
\nabla_Y^+\nabla_Z^+\nabla_W^+\psi_X&=&\int_0^1\prod d\alpha_i\delta
(\sum\alpha_i-1)\mbox{Tr}\left(\rho_X^{\alpha_1}Y\rho_X^{\alpha_2}Z\rho_X
^{\alpha_3}W\right)_c\nonumber\\
&:=&t_X(Y,Z,W).\label{torsion}
\end{eqnarray}
Here, $\nabla_Y^+$ denotes the Gateaux derivative of $\psi_X$ in the direction
$Y\in{\cal B}({\cal H})$, $\hat{Y}=Y-\rho_X.Y$ is the centred observable
(the socalled ``score''), $g_X(Y,Z)$ is the Bogoliubov-Kubo-Mori ({\em BKM})
metric,
and ${\rm Tr}(\ldots)_c$ denotes the cumulant of third order.
We adopt the convention throughout this paper that if $\lambda\in[0,1]$ then
$\lambda^\prime=1-\lambda$.
The torsion, $t$, measures the failure of $g$ to be invariant under
$(+1)$-parallel transport.

In \cite{RFS1,RFS2,RFS3} we began a study of the unbounded case. Let
${\cal C}_p,\hspace{.1in}0<p<1$ denote the set of compact maps $A:{\cal H}
\rightarrow{\cal H}$ such that $|A|^p\in{\cal C}_1$, and put
\begin{equation}
{\cal C}_{<1}:=\bigcup_{0<p<1}{\cal C}_p.
\end{equation}
It is known \cite{Pietsch} that if $p<q$ then ${\cal C}_p\subset{\cal C}
_q$ and that ${\cal C}_p$ is a complete linear space with quasinorm
\begin{equation}
\|A\|_p:=[{\rm Tr}\left(A^*A\right)^{p/2}]^{1/p}.
\end{equation}
In \cite{RFS2} we took the underlying set of the quantum information
manifold to be
\begin{equation}
{\cal M}={\cal C}_{<1}\cap\Sigma.
\end{equation}
${\cal M}$ has a natural affine structure, coming from the linear structure
of each ${\cal C}_p$; it is called the $(-1)$-affine structure.
Thus if $\rho_i\in{\cal C}_{p_i}\cap\Sigma, i=1,2$
let $p=\max\{p_1,p_2\}$; then $\rho_i\in{\cal C}_p\cap\Sigma, i=1,2$. We
define $\lambda\rho_1+\lambda^\prime\rho_2, 0\leq\lambda\leq 1$ as the
usual sum of operators in ${\cal C}_p$. We see that ${\cal M}$ is
$(-1)$-convex.

There is a quantum analogue of the classical Orlicz space $L\log L$: let
\begin{equation}
{\cal C}_1\log{\cal C}_1:=\{\rho\in{\cal C}_1:S(\rho):=-\sum_{i=1}^\infty
\lambda_i\log\lambda_i<\infty\}
\end{equation}
where $\{\lambda_i\}$ are the singular numbers of $\rho$. Thus the set of
normal states of finite entropy is ${\cal C}_1\log{\cal C}_1\cap\Sigma$. It
is easy to show that ${\cal C}_{<1}\subset{\cal C}_1\log{\cal C}_1$.

The topology given to ${\cal M}$ in \cite{RFS2} is not that induced on it as a
subset of ${\cal C}_1$. Following an idea in \cite{Pistone} we constructed
a hood of $\rho_0\in{\cal M}$ consisting of all $\rho\in{\cal M}$
which are `connected to $\rho_0$ by a one-parameter exponential family
of states'. In \cite{RFS2} this is the family
\begin{equation}
\rho_{\lambda X}=Z_{\lambda}^{-1}e^{-(H_0+\lambda X+c(\lambda))}\in{\cal M},
\hspace{.3in}0\leq\lambda\leq 1,
\end{equation}
where $X$ is a symmetric quadratic form defined on ${\cal D}(H_0^{1/2})$
and form-bounded relative to the energy form
\begin{equation}
q_0(\phi,\psi):=\langle H_0^{1/2}\phi,H_0^{1/2}\psi\rangle.
\end{equation}
The state $\rho_{\lambda X}$ does not depend on the choice of $c(\lambda)$
and we do this so that $H:=H_0+\lambda X+c(\lambda)\geq I$. We write
$R=R_X:=H^{-1}$ for the inverse of $H$ at $\lambda=1$.
If $\rho_0\in{\cal C}_{\beta_0}$, to make sense we need that the $q_0$-bound
of $X$ be less than $1-\beta_0$. We then showed that $\rho_X\in{\cal C}_
{\beta_X}$ for some $\beta_X<1$. The set of $q_0$-bounded quadratic forms
$X$ is a Banach space ${\cal T}_0$ with norm
\begin{equation}
\|X\|_0:=\|R_0^{1/2}XR_0^{1/2}\|,
\end{equation}
where $\|\bullet\|$ is the operator norm for ${\cal H}$, and $R_0=H_0^{-1}$.
We showed that sufficient for $\rho_X\in{\cal M}$ is that $\|X\|_0
<1-\beta_0$, and that $Z_X$ is a Lipschitz function of $X$; however, we
could not show that the first moment ${\rm Tr}(\rho_0X)$ is finite. We
introduced the regularised mean of $X\in{\cal T}_0$,
\begin{equation}
\rho_0.X:={\rm Tr}[\rho_0^\lambda X\rho_0^{\lambda^\prime}]
\end{equation}
and showed that it is finite and independent of $\lambda\in(0,1)$.

The hood of $\rho_0$ defined by $\{\rho_X:\|X\|_0<1-\beta_0\}$
can be mapped bijectively onto the open ball of radius $1-\beta_0$ in the
Banach space
\begin{equation}
\hat{\cal T}_0:=\{X\in{\cal T}_0:\rho_0.X=0\};
\end{equation}
thus, $\hat{\cal T}_0$ consists of score variables. This map makes the
hood into a manifold modelled on $\hat{\cal T}_0$. The linear
structure of $\hat{\cal T}_0$ induces a local affine structure on ${\cal M}$,
called the $(+1)$-affine structure. To any point $\rho_X$, we may similarly
define a hood of states of $\rho_X$ to be
\begin{equation}
\left\{\rho_Y:\|Y\|_X:=\|R_X^{1/2}YR_X^{1/2}\|<a\leq 1-\beta_X\right\},
\end{equation}
which is in bijection with an open ball in the Banach space
\begin{equation}
\hat{\cal T}_X:=\{Y:\|Y\|_X<\infty\mbox{ and }\rho_X.Y=0\}.
\end{equation}
We showed that on the overlap region, the norms $\|\bullet\|_0$ and $\|
\bullet\|_X$ are equivalent norms; the $(+1)$-affine structures on the
region of overlap are also the same. Thus we have covered ${\cal M}$
by a Banach manifold
with a local affine structure. We showed that the connected component
containing $\rho_0$ is $(+1)$-convex.

In the present paper we consider the problem of furnishing ${\cal M}$
with a Riemannian metric. As shown by Petz \cite{Petz2}, in the
finite-dimensional case there are several candidates, but for
geometric reasons we choose the {\em BKM} metric. Indeed, the studies of
Nagaoka \cite{Nagaoka} show that $+1$ and $-1$-affine structures are not
dual relative to the `Uhlmann' metric, (the real part of ${\rm Tr}\,
\rho X^*Y)$). They {\em are} dual, however, relative to the {\em BKM}-metric.
Since this is the second derivative of $\psi$, in infinite dimensions
we must impose more regularity than we have so far. Apart from having this
`Amari' duality, similar to the classical Fisher-Rao metric, the
{\em BKM}-metric is the one entering Kubo's theory of linear response,
as well as the one arising in quantum statistical dynamics.

In this paper we assume that $X$ is an operator which is $H_0$-bounded.
We then say that the tangent direction $X$ is analytic. We show in \S(2)
that this leads to the Fr\'{e}chet differentiability of $\psi_0$ in the
analytic direction $X$, with Lipschitz continuous derivative
\begin{equation}
\nabla_X\psi_0=\rho_0.X,
\end{equation}
and that $\rho_0 X$ is of trace class. This means that the regularisation
of the mean is not necessary.
We show that a metric $g$ can be defined by eq.~(\ref{met}).

In \S(3) we show that the $n$-point Kubo cumulant is finite, and that
for small $|\lambda|$ the Taylor series for $Z_{\lambda V}$ converges
if $V$ is an analytic direction. We show that when we furnish ${\cal M}$
with the analytic topology, it becomes a real analytic Banach manifold.

\section{$H_0$-bounded perturbations}
Let $H_0\geq I$ be a selfadjoint operator with domain ${\cal D}(H_0)$
and quadratic form $q_0$. A necessary and sufficient condition for a
symmetric operator $X:{\cal D}(H_0)\rightarrow{\cal H}$ to be $H_0$-bounded
is that $R_0X$ be bounded \cite{RS}. One can show that
\begin{equation}
\|X\|_0:=\|R_0^{1/2}XR_0^{1/2}\|\leq\|R_0X\|\label{stronger}
\end{equation}
It follows that if $X$ is $H_0$-bounded, then its
form is $q_0$-bounded. Suppose now that $\rho_0:=Z_0^{-1}e^{-H_0}\in{\cal C}
_{\beta_0}$, $\beta_0<1$, and that $\|R_0X\|<1-\beta_0$. Then by
eq.~(\ref{stronger}), $\|X\|_0<1-\beta_0$ and so by \cite{RFS2} the
perturbed state
\begin{equation}
\rho_X=Z_X^{-1}e^{-H_X}\in{\cal C}_{\beta_X}\subset{\cal C}_{<1}
\end{equation}
lies in a hood of $\rho_0$ in the topology induced by ${\cal T}_0$.
Let us define a stronger topology than this, by taking hoods
of $\rho_0$ to be
\begin{equation}
\{\rho_X:\|X\|_\omega:=\|R_0X\|<a\leq1-\beta_0\}.
\end{equation}
We call this the analytic or $\omega$-topology, and $\|\bullet\|_\omega$
the $\omega$-norm. Similarly, if $\rho_X$ is in the
analytic hood of $\rho_0$, and $Y$ is $H_X$-bounded with $\|R_XY\|<1-
\beta_X$, then $\rho_{X+Y}\in{\cal M}$ and the analytic hoods of $\rho_X$
can be defined by
\begin{equation}
\{\rho_{X+Y}:\|R_XY\|<a\leq1-\beta_X\}.
\end{equation}
Let $\hat{\cal T}_\omega(\rho_X):=\{Y:\|YR_X\|<\infty,\hspace{.2in}\rho_X.Y=0\}.$
Then the construction of the $\omega$-hoods covers ${\cal M}$ by sets
homeomorphic to balls in the Banach spaces $\hat{\cal T}_\omega(\rho)$; these
have mutually equivalent norms on the overlaps of the hoods. To see this,
suppose that $Y$ is in the hood of $\rho_0$ and also of $\rho_X$. The charts
around $\rho_0$ and $\rho_X$ take $\rho_Y$ to balls in the Banach spaces
$\hat{\cal T}_\omega(\rho_0)$ and $\hat{\cal T}_\omega(\rho_X)$, with norms
$\|R_0Y\|$ and $\|R_XY\|$. These are equivalent, since
\begin{equation}
m\|YR_0\|\leq\|YR_X\|\leq M\|R_0Y\|\hspace{.5in}\mbox{for all }Y,
\end{equation}
where $m=\|(H_X+I)R_0\|^{-1}$ and $M=\|(H_0+I)R_X\|$.
Thus ${\cal M}$ is made into a Banach manifold, called the $\omega$
quantum information manifold. This name will be justified in the next
section. It is easy to see that the local
$(+1)$-affine structure coming from the linear structure of $\hat{\cal T}
_\omega(\rho_X)$ is compatible with that given in \cite{RFS2}.

The following lemma will be needed later. It is clear that $R_{X+Y}$ maps
${\cal H}$ into ${\cal D}(H_0)={\cal D}(H_X)={\cal D}(H_{X+Y})$. It follows
from lemma 6, \cite{RFS2} that $H_XR_{X+Y}$ is bounded. We find
\begin{equation}
(I+YR_X)(H_XR_{X+Y})=H_XR_{X+Y}+YR_{X+Y}=I.
\end{equation}
Hence $H_XR_{X+Y}=(I+YR_X)^{-1}$, and we get if $\|YR_X\|<1$,
\begin{equation}
\|H_XR_{X+Y}\|\leq(1-\|YR_X\|)^{-1}.
\end{equation}
From this we see
\begin{lemma}\label{two}
If $a<1$ and $X$ is $H_0$-small, then $\|H_XR_{X+Y}\|$ is bounded in the set
$\{Y:\|YR_X\|\leq a\}$.\hspace{\fill}$\Box$
\end{lemma}
Now suppose that $X$ perturbs $H_0$ and $Y$ perturbs $H_X$, as above. Then
\begin{theorem}
If $1-2\delta>\beta_X$, then $\rho_X^{\delta^\prime}Y$ has a trace-class
extension.
\end{theorem}
Proof. Choose $\delta>0$ so that $\rho_X^{1-2\delta}\in{\cal M}$. Then
\begin{eqnarray*}
\|\rho_X^{1-\delta}Y\|_1&=&\|Z_X^{\delta-1}e^{-(1-\delta)H_X}Y\|_1\\
&\leq&C\|\rho_X^{1-2\delta}\|_1\|e^{-\delta H_X}H_X\|_\infty\|R_XY\|_\infty\\
&<&\infty.
\end{eqnarray*}
\hspace*{\fill}$\Box$\\
Corollary. The regularised mean coincides with the true mean:
\begin{equation}
{\rm Tr}(\rho_X^{\lambda}Y\rho_X^{1-\lambda})={\rm Tr}(\rho_XY),
\hspace{.2in}\mbox{ for all }\lambda\in(0,1).
\end{equation}
For, we can write $\rho_XY=\rho_X^\delta\rho_X^{\delta^\prime}Y$ and use
the cyclicity of the trace.
\begin{theorem}
In the analytic manifold, $\psi_X:=\log Z_X$ is $\nabla_Y^+$ Fr\'{e}chet
differentiable, and
\begin{equation}
\nabla_Y^+\psi_X=\rho_X.Y
\end{equation}
\end{theorem}
Proof. By two applications of Duhamel's formula \cite{RFS2}, Th. 9, we get
for the difference quotient,
\begin{eqnarray*}
& &\lambda^{-1}{\rm Tr}\left[e^{-(H_X+\lambda Y)}-e^{-H_X}\right]-{\rm Tr}
\left[e^{-H_X}Y\right]=\\
&=&{\rm Tr}\int_0^1\alpha d\alpha\int_0^1d\beta
\left\{e^{-\alpha\beta(H_X+\lambda Y)}\lambda Ye^{-\alpha\beta^\prime H_X}
Ye^{-\alpha^\prime H_X}\right\}.
\end{eqnarray*}
We can put the trace inside the integral, by Fubini's theorem, if we can
show that by doing so we get an absolutely convergent integral of the
trace norm. The right hand side is then obviously $O(\lambda)$, with
constant equal to the trace of an operator of the form
\begin{equation}
C=\int\int\int_0^1\rho_1^{\alpha_1}Y\rho_2^{\alpha_2}Z\rho_3^{\alpha_3}
\prod_{i=1}^3d\alpha_i\delta\left(\sum_{i=1}^3\alpha_i-1\right),
\end{equation}
which is the quantum version of the integral remainder in Taylor's theorem.

We divide the region of integration into three (overlapping) parts; in
region $i$, $\alpha_i\geq 1/3, (i=1,2,3)$. The estimate is similar in each
case. The hardest case is when the variable larger than 1/3, say $\alpha_1$,
involves the state $\rho_1=Z_\lambda^{-1}e^{-(H_X+\lambda X)}$, as this
requires an estimate independent of $\lambda$. We therefore do this case
in detail. So take $\alpha_1\geq 1/3$, and $\alpha_2,\alpha_3>0$, and
$H_i\geq I$. Put $R_i=H_i^{-1}$. We now
show that for small $\delta>0$, $\rho_1^{\alpha_1-\delta}Y\rho_2^{\alpha_2}
Z\rho_3^{\alpha_3}$ is of trace class. Indeed, by H\"{o}lder,
\begin{eqnarray*}
\|\rho_1^{\alpha_1-\delta}Y\rho_2^{\alpha_2}Z\rho_3^{\alpha_3}\|_1&\leq&
\|\rho_1^{\alpha_1-2\delta}\|_{1/\alpha_1}\|Y\rho_1^\delta\|_\infty\|\rho_2
^{\alpha_2}\|_{1/\alpha_2}\\
& &\|Z\rho_3^\gamma\|_\infty\|\rho_3^{\alpha_3-\gamma}\|_{1/\alpha_3}\\
&\leq&\|\rho_1^{1-2\delta/\alpha_1}\|_1^{\alpha_1}\|YR_1\|_\infty
\|H_1\rho_1^\delta\|_\infty\\
& &\|ZR_3\|_\infty\|H_3\rho_3^\gamma\|_\infty\|\rho_3^{1-\gamma/
\alpha_3}\|_1^{\alpha_3}.
\end{eqnarray*}
If $2\delta<\beta_1^\prime\alpha_1$ and $\gamma<\alpha_3\beta_3^\prime$,
then all
the norms are finite; here, the $\beta_i$ are such that $\rho_i\in{\cal C}_
{\beta_i}$. The size of $\gamma$ depends on $\alpha_3$, but in the region
$\alpha_1\geq 1/3$, $\delta$ can be chosen independent of $\alpha_i$.
Now take $2\delta<\beta_1^\prime/3$. Then for
each $\alpha_1\geq 1/3$, $\alpha_2>0$, $\alpha_3>0$ we can take a bit,
$\rho_1^\delta$, of the dominant factor to the right end. We get, using
$\sum\alpha_i=1$ and H\"{o}lder,
\begin{eqnarray*}
\|C\|_1&=&\|\rho_1^{\alpha_1-2\delta}(\rho_1^\delta Y)(\rho_2^{\alpha_2})
(ZR_3)\rho_3^{\alpha_3}(H_3\rho_1^\delta)\|_1\\
&\leq&\|\rho_1^{\alpha_1-2\delta}\|_{1/\alpha_1}\|\rho_1^\delta Y\|_\infty\|
\rho_2^{\alpha_2}\|_{1/\alpha_2}^{\alpha_2}\|ZR_3\|_\infty\\
& &\|\rho_3^{\alpha_3}\|_{1/\alpha_3}^{\alpha_3}\|H_3R_1\|_\infty\|H_1\rho_1
^\delta\|_\infty.
\end{eqnarray*}
The first norm is bounded by
\[ \|\rho_1^{1-2\delta/\alpha_1}\|_1^{\alpha_1}\leq \|\rho_1^{1-6\delta}\|_1
<\infty\]
as $1-6\delta>\beta_1$.
The second factor,
\[\|\rho_1^\delta Y\|_\infty\leq \|\rho_1^\delta H_1\|_\infty\|R_1Y\|_
\infty\]
is finite, independent of $\lambda$, by the spectral theorem,
and by lemma~(\ref{two}). The same can be said of the factors
$\|H_1\rho_1^\delta\|_\infty$ and $\|H_3R_1\|_\infty$. The factors involving
$\rho_2$ and $\rho_3$ are unity, as they are states.
Hence the trace norm of $C$ is bounded (in the region 1) independent of
$\alpha$ and $\lambda$, and its integral is bounded. The other regions, 2
and 3, are treated similarly.\hspace{\fill}$\Box$\\
Corollary. The {\em BKM} metric
\begin{equation}
g_X(Y,Z)=\int_0^1d\alpha{\rm Tr}\left(\rho_X^\alpha Y\rho_X^
{1-\alpha}Z\right)
\end{equation}
is finite if $Y,Z$ are $\omega$-directions. Moreover, $\psi_X$ is Fr\'{e}chet
differentiable in the $Y$ direction in the analytic manifold, with
derivative $\rho_X.Y$; for
\begin{equation}
|\rho_X.Y|=|{\rm Tr}[\rho_X^{1-\delta}(\rho_XH_X)(R_XY)]|\leq C\|R_XY\|
\end{equation}
is continuous in the $\omega$-norm

\section{The analyticity of the free energy}
In this section, we show that the free energy $\psi_{\lambda X}$ is an
analytic function in a hood of $\lambda=0$ if $X$ is an $\omega$-direction. 
Suppose $\rho=Z^{-1}\exp(-H)\in{\cal C}_\beta\in{\cal M}$, and $V_j$
are $H$-bounded, $j=1,\ldots,n$. As before, we assume that $H\geq 1$ and
write $R=H^{-1}$.

We start with the $n$-point function
\[M_n:=Z_0{\rm Tr}\int_0^1d\alpha_1\int_0^1d\alpha_2\ldots\int_0^1d\alpha_{n-1}\left[
\rho^{\alpha_1}V_1\rho^{\alpha_2}V_2\ldots\rho^{\alpha_n}V_n\right].\]
Here, $\alpha_n=1-\alpha_1-\ldots-\alpha_{n-1}$.
Write this trace as
\begin{eqnarray*}
& &\left[\rho^{\alpha_1\beta}\right]\left[H^{1-\delta_n+\delta_1}
\rho^{(1-\beta)\alpha_1}\right]
\left[R^{\delta_1}V_1R^{1-\delta_1}\right]\left[
\rho^{\alpha_2\beta}\right]\left[H^{1-\delta_1+
\delta_2}\rho^{(1-\beta)\alpha_2}\right]\\
& &\left[R^{\delta_2}V_2R^{(1-\delta_2)}\right]\ldots
\left[\rho^{\alpha_n\beta}\right]\left[H^{1-\delta_{n-1}+\delta_n}
\rho^{(1-\beta)\alpha_n}\right]\left[R^{\delta_n}V_nR^{1-\delta_n}\right],
\end{eqnarray*}
where $\delta_j\in(0,1)$ are yet to be chosen.
This is the product of $n$ factors of the form $\left[\rho^
{\alpha_j\beta}\right]$, $n$ factors of the form 
$\left[H^{1-\delta_{j-1}+\delta_j}\rho^{(1-\beta)\alpha_j}\right]$,
where $\delta_0$ means $\delta_n$,
and $n$ factors of the form $\left[R^{\delta_j}V_jR^{1-\delta_j}\right]$.
By interpolation, the last type is bounded in operator norm by \cite{RS}
\[ \|R^{\delta_j}V_jR^{1-\delta_j}\|\leq \|RV_j\|=\|V_j\|_\omega.\]
This bound is independent of $\alpha$. The factors $\rho^{\alpha_j\beta}$
are bounded in trace-norm by the H\"{o}lder inequality
\[\|A_1\ldots A_n\|_1\leq \|A_1\|_{p_1}^{1/p_1}\ldots\|A_n\|_{p_n}^{1/p_n}.\]
Since $\sum\alpha_j=1$ we may put $p_j=1/\alpha_j$ to get
\begin{eqnarray*}
\|\left[\rho^{\alpha_1\beta}\right]&\ldots&\left[\rho^{\alpha_n\beta}\right]
\|_1 \leq \|\rho^\beta\|_1^{\alpha_1}\ldots\|\rho^\beta\|_1^{\alpha_n}\\
&=&\|\rho^\beta\|_1<\infty.
\end{eqnarray*}
This bound is also independent of the $\alpha's$.
We bound the remaining factors in operator norm.
By the spectral theorem we have the bound
\begin{eqnarray}
& &\|\rho^{(1-\beta)\alpha_j}H^{1-\delta_{j-1}+\delta_j}\|_\infty=Z^
{-\alpha_j(1-\beta)}\sup_{x\geq 1}\{e^{-(1-\beta)\alpha_jx}x^{1-\delta_{j-1}+
\delta_j}\}\\
& &\leq Z^{-\alpha_j(1-\beta)}\left(\frac{1-\delta_{j-1}+\delta_j}
{(1-\beta)\alpha_j}\right)^{1-\delta_{j-1}+\delta_j}e^{-(1-\delta_{j-1}+
\delta_j)}.\label{bound}
\end{eqnarray}
We have to integrate $\alpha_j^{-(1-\delta_{j-1}+\delta_j)}d\alpha_j$; the
region of integration is the union of $n$ (overlapping) regions
$S_j:=\{\alpha: \alpha_j\geq 1/n\}$. We treat each in a similar manner. For
$S_n$ we can ensure integrability
at $\alpha_j=0$, $j=1,\ldots,n-1$ if we choose $\delta_j$ so that
$1-\delta_{j-1}+\delta_j<1$, that is, $\delta_j<\delta_{j-1}$.
So we choose $\delta_n=\delta_0>\delta_1>\delta_2>\ldots>\delta_{n-1}$.
To ensure that $\delta_j\in[0,1]$ it is enough to choose $\delta_n=1,
\;\delta_1=1-1/n,\;\ldots,\delta_{n-1}=1/n$.
Then each of the $n-1$ integrals for $j=1,\ldots n-1$,
\[\int_0^1d\alpha_j\alpha_j^{-(1-\delta_{j-1}+\delta_j)}=(\delta_{j-1}-
\delta_j)^{-1}\]
is equal to $n$, giving a factor $n^{n-1}$. The remaining $\alpha$ factor in
the region $S_n$ is $\alpha_n^{-(1-\delta_{n-1}+\delta_n)}\leq n^2$. We get
the same bound in the other regions, $S_j$ $j=1,\ldots n-1$, so we get
the bound $n^3n^{n-1}=n^2n^n$. The other factors in the estimate
(\ref{bound}) can be bounded independent of $\alpha$:
\begin{eqnarray*}
& &\prod_{j=1}^{n}Z^{-\alpha_j(1-\beta)}(1-\delta_{j-1}+\delta_j)^
{1-\delta_{j-1}+\delta_j}(1-\beta)^{-(1-\delta_{j-1}+\delta_j)}
e^{-(1-\delta_{j-1}+\delta_j)}\\
& &\leq 4Z^{-(1-\beta)}(1-\beta)^{-n}e^{-n}
\end{eqnarray*}
since $(1-\delta_{j-1}+\delta_j)<1$ except for one term, when it is less
than $2$. Thus we get for the $n$-point function the bound
\[4\|\rho^\beta\|_1Z^{-\beta}n^2n^ne^{-n}\prod_j\|V_j/(1-\beta)\|_\omega.\]
It follows that the partition function, $Z_{\lambda V}$, has a convergent
Taylor series $\sum\lambda^n M_n/n!$x if $|\lambda|\,\|V\|_\omega<1-\beta$.
Recall that this is also the condition that $\rho_{\lambda V}$ lie in
${\cal M}$ in an $\omega$-hood of $\rho$. Since $Z_{\lambda V}>0$, it
follows that the free energy, $\psi:=\log Z_{\lambda V}$, is real
analytic in $\lambda$ in the same interval.

\section{Conclusion}
Suppose that $\beta<1$. We have seen that the condition that the
perturbation of the Hamiltonian $H\geq I$
of a state $\rho\in{\cal C}_\beta$ by a small enough symmetric operator $V$
leads to another state $\rho_V\in{\cal M}$. We defined a topology on
${\cal M}$ in a hood of $\rho$ using the norm $\|V\|_\omega(\rho)=\|RV\|$,
where $RH=I$; this hood ${\cal U}$ is set of states $\rho_V$ coming from
the ball of operators $V$ with $\|V\|_\omega<1-\beta$. This is
finer than the topology defined by the norm $\|V\|=\|R^{1/2}VR^{1/2}\|$
which is finite for all form-bounded perturbations.
We saw that the Taylor series for the partition function converges in the
same hood of $\rho$. The $\omega$-norms associated
with different points of ${\cal M}$ are equivalent on the overlap of
two hoods, so together we have a Banach manifold modelled on the Banach
spaces of $H$-bounded symmetric operators. The question arises, is there an
analytic structure of this manifold? In fact, an analytic structure is
deemed to be provided if we specify the ring of germs of
analytic functions at each point. Let us say that a map $\psi:{\cal U}
\rightarrow{\bf C}$ is $+1$-analytic if in ${\cal U}$ it is
infinitely often Fr\'{e}chet differentiable and $\psi(\rho_{\lambda V})$
has a convergent Taylor expansion in $\lambda$, for all $\omega$-directions
$V$ in the tangent space. Because of the equivalence of norms, this
concept is coordinate-free. Thus an analytic structure for ${\cal M}$
has been specified. In particular, the mixture coordinates $\eta_X=\rho.X$
are analytic, as they are derivatives of the free energy.
In his theory of expansionals, Araki \cite{Araki} showed that
the free energy $\psi_X$ is an entire function in the Banach space
${\cal B}({\cal H})$ if the perturbation $X$ is bounded,
in the more general context of Tomita-Takesaki theory. With unbounded
perturbations we cannot hope for entire functions; for we need only take
$V=-H_0$ to hit a singularity in $\psi_X$. However, it is likely that some
analyticity remains, even in the thermodynamic limit, for relatively bounded
perturbations.

Compared with \cite{RFS1}, we have improved the result in several ways.
We show analyticity instead of twice-differentiability; we have dropped
the commutator condition altogether; the manifold is infinite dimensional
instead of finite-dimensional; and we have enlarged the class of states
to ${\cal C}_{<1}$. Further results are obtained in \cite{MRGRFS}.

\vspace{.1in}
\noindent{\bf Acknowledgements} This continues work started at CNRS Luminy,
Marseille; I thank P. Combe for arranging the visit, and CNRS for
financial support. Thanks are due to P. Combe, G. Burdet and
H. Nencka for useful discussions.


\begin{thebibliography}{99}
\bibitem{Amari} S.-i. Amari, {\em Differential Geometric Methods in
Statistics,} {\bf Lecture Notes in Statistics}, {\bf 28}, 1985.
Springer-Verlag.
\bibitem{Araki} Araki, H., Publ. R. I. M. S. (Kyoto), {\bf 9}, 165-209, 1968.
\bibitem{Balian} Balian, R., Y. Alhassid, H. Reinhardt, {\em Dissipation
in many-body systems: a geometrical approach based on information theory},
Physics Reports, {\bf 131}, 1-146, 1986. North Holland.
\bibitem{Bogoliubov} Bogoliubov, N. N., Phys. Abh. Sov. Union, {\bf 1}, 229-,
1962.
\bibitem{Gibilisco} Gibilisco, P., and G. Pistone, {\em Connections on
nonparametric statistical manifolds by Orlicz space geometry},
Infinite-dimensional Analysis, Quantum Probability and Related Topics,
{\bf 1}, 325-347, 1998.
\bibitem{MRGRFS} Grasselli, M. R., and R. F. Streater, {\em The quantum
information manifold for epsilon-bounded forms}, math-ph/9910031.
\bibitem{Hasagawa} Hasagawa, H. Reports on Math. Phys., {\bf 33}, 87, 1993.
\bibitem{Hasagawa2} Hasagawa, H. {\em Noncommutative Extension of the Information
Geometry}, in {\bf Quantum Communication and Measurement}, Eds V. P.
Balavkin, O. Hirota and R. L. Hudson, 1995, Plenum Press, pp 327-337,
\bibitem{Kass} Kass, R. A. and P. W. Vos, ``Geometric Foundations of
Asymptotic Inference'', Wiley, NY 1997.
\bibitem{Kubo} Kubo, R. Reports on Progress in Physics, {\bf 29}, 255-284,
1966.
\bibitem{Lesniewski} Lesniewski, A., and M. B. Ruskai, {\em Relative Entropy
and Monotone Riemannian Metrics on Noncommutative Probability Spaces},
preprint.
\bibitem{Matsubara} Matsubara, T., Prog. Theor. Phys.,{\bf 14}, 351-, 1955.
\bibitem{Mori} Mori, H., Prog. Theor. Phys., {\bf 33}, 423-, 1965.
\bibitem{Nagaoka} Nagaoka, H., pp 449-452 in {\bf Quantum Communication
and Measurement}, Eds V. P. Belavkin, O. Hirota and R. L. Hudson, Plenum
Press, New York, 1995.
\bibitem{RFS3} H. Nencka and R. F. Streater, {\em Information Geometry for
Some Lie Algebras}, to appear in Infinite Dimensional Analysis and
Quantum Probability, World Scientific.
\bibitem{Petz} Petz, D., and Toth, G. Lett. Math. Phys., {\bf 27}, 205-216,
1993.
\bibitem{Petz2} Petz, D., {\em Monotone Metrics on Matrix Spaces}, Linear Alg.
and Appl., {\bf 244}, 81-96, 1996.
\bibitem{Petz3} Petz, D. and C. Sudar, {\em Geometries of Quantum States},
Journ. Math. Phys., {\bf 37}, 2662-73, 1996.
\bibitem{Pietsch} Pietsch, A. {\bf Nuclear Locally Convex Spaces},
Springer-Verlag, 1972.
\bibitem{Pistone} Pistone, G. and C. Sempi, Annals of Statistics, {\bf 23},
1543-1561, 1995.
\bibitem{RS} Reed, M. and B. Simon, {\bf Methods of Modern Mathematical
Physics}, Academic Press, Vol. 2, 1975.
\bibitem{Roepstorff} Roepstorff, G. Commun. Math. Phys.,{\bf 46}, 253-262
1976.
\bibitem{RFS1} R. F. Streater, {\em Information Geometry and Reduced
Quantum Description}, Reports on Mathematical Physics, {\bf 38}, 419-436
1996.
\bibitem{RFS2} R. F. Streater, {\em The Information Manifold for Relatively
Bounded Potentials}, to appear in the Bogoliubov $90^{\rm th}$ Birthday
Memorial Volume. Ed A. A. Slavnov. math-ph/9910035

\end{thebibliography}
\end{document}